\documentstyle[prl,aps,epsf,epsfig]{revtex}

\begin{document}
\draft

\title{Master Operators Govern Multifractality in Percolation
}
\author{O. Stenull and H. K. Janssen
}
\address{
Institut f\"{u}r Theoretische Physik 
III\\Heinrich-Heine-Universit\"{a}t\\Universit\"{a}tsstra{\ss}e 1\\40225 
D\"{u}sseldorf, Germany
}
\date{\today}
\maketitle

\begin{abstract}
Using renormalization group methods we study multifractality in percolation at the instance 
of noisy random resistor networks. We introduce the concept of master operators. The 
multifractal moments of the current distribution (which are proportional to the noise 
cumulants $C_R^{(l)} \left( x, x^\prime \right)$ of the resistance between two sites $x$ and 
$x^\prime$ located on the same cluster) are related to such master operators. The scaling 
behavior of the multifractal moments is governed exclusively by the master operators, even 
though a myriad of servant operators is involved in the renormalization procedure. We 
calculate the family of multifractal exponents $\left\{ \psi_l \right\}$ for the scaling 
behavior of the noise cumulants, $C_R^{(l)} \left( x, x^\prime \right) \sim \left| x - 
x^\prime \right|^{\psi_l /\nu}$, where $\nu$ is the correlation length exponent for 
percolation, to two-loop order. 
\end{abstract}

Percolation\cite{bunde_havlin_91_etc} is a leading paradigm for disorder. Though it 
represents the simplest model of a disordered system, it has many applications, e.g., 
polymerization, porous and amorphous materials, thin films, spreading of epidemics etc. In 
particular the transport properties of percolation clusters have gained a vast amount of 
interest over the last decades. Random resistor networks (RRN) are the most prominent model 
for transport on percolation clusters. Here we discuss RRN in the context of  
multifractality\cite{evertsz_mandelbrot_92}. We propose the concept of master operators 
('operator' in the sense of 'composite field'). Each moment of the multifractal measure (the 
square of the bond currents) has a master operator as field theoretic counterpart. These 
master operators are highly and dangerously irrelevant in the renormalization group sense. 
Therefore, each master operator needs in general a myriad of other irrelevant operators for 
renormalization. However, the renormalization of these servant operators does not induce 
their master. It follows, that the servant operators can be neglected in determining the 
scaling index of their master operator, i.e., one is spared the computation and 
diagonalization of gigantic renormalization matrices. 

Systems where multifractality has been observed include, beside RRN, turbulence, diffusion 
near fractals, electrons in disordered media, polymers in disordered media, random 
ferromagnets, and chaotic dissipative systems. Multifractality in RRN is exhibited by the 
moments of the current distribution measured between two connected terminals $x$ and 
$x^\prime$. The multifractal moments of the current distribution are related to the noise 
cumulants $C^{(n)}_R \left( x, x^\prime \right)$ of the resistance by Cohn's 
theorem\cite{cohn_50}. At criticality the noise cumulants scale as $C_R^{(n)} \left( x, 
x^\prime \right) \sim \left| x - x^\prime \right|^{-x_n}$, where the $x_n$ constitute an 
infinite set of exponents which are not related to each other in a linear way as commonly 
occurs in critical phenomena under the name of gap scaling. The multifractality manifests 
itself in the nonlinear dependence of 
$x_n$ on $n$. The existence of the set $\left\{ x_n \right\}$ was proposed by Rammal 
{\em et al}\cite{rammal_etal_85}. A set of exponents $\left\{ \zeta_{2n} \right\}$ 
equivalent to $\left\{ - x_n \nu \right\}$, where $\nu$ is the correlation length 
exponent for percolation, was also proposed by Arcangelis {\em et 
al.}\cite{arcangelis_etal_85}.

The field theoretic description of multifractality in RRN was pioneered by Park, Harris 
and Lubensky (PHL)\cite{park_harris_lubensky_87}. Based on an approach by 
Stephen\cite{stephen_78} they formulated a $D \hspace{-1mm}\times \hspace{-1mm}E$-fold 
replicated Hamiltonian to study the effects of noise in RRN. The contributions to the 
Hamiltonian leading to multifractal behavior contain powers of replica space gradients 
analog to powers of real space gradients, which were accounted for as an origin of 
multifractality by Duplantier and Ludwig\cite{duplantier_ludwig_91}. PHL introduced a 
set of exponents $\left\{ \psi_n \right\}$ identical to the set $\left\{ - x_n \nu 
\right\}$ and calculated it to first order in $\epsilon = 6-d$, where $d$ denotes the 
spatial dimension.  

Consider a $d$-dimensional lattice, where bonds between nearest neighboring sites $i$ 
and $j$ are randomly occupied with probability $p$ or empty with probability $1-p$. Each 
occupied bond $b = \langle i,j \rangle$ has a conductance $\sigma_b$. Unoccupied bonds have 
conductance zero. The bonds obey Ohm's law $\sigma_{i,j} \left( V_j - V_i \right) = I_{i,j}$, 
where $I_{i,j}$ is the current flowing through the bond from $j$ to $i$ and $V_i$ is the 
potential at site $i$. The $\sigma_b$ are equally and independently distributed random 
variables with distribution function $f$, mean $\overline{\sigma}$, and higher cumulants 
$\Delta^{(n\geq 2)}$ satisfying $\Delta^{(n)} \ll \overline{\sigma}^n$. The noise average is 
denoted by $\left\{ \cdots \right\}_f = \int \prod_b d \sigma_b f \left( \sigma_b \right) 
\cdots$ and its $n$th cumulant by $\left\{ \cdots^n \right\}^{(c)}_f$. Both kinds of 
disorder, the random dilution of the lattice and the noise, influence the statistical 
properties of the resistance $R (x ,x^\prime )$ of the backbone between two sites $x$ and 
$x^\prime$. They are reflected by the 
noise cumulants 
\begin{eqnarray}
\label{defCumulant}
C_R^{(n)}(x ,x^\prime) = \left\langle \chi (x ,x^\prime) \left\{ R (x 
,x^\prime )^n \right\}_f^{(c)} \right\rangle_C / \left\langle \chi (x ,x^\prime) 
\right\rangle_C \ , 
\end{eqnarray}
where $\langle \cdots \rangle_C$ denotes the average over the configurations $C$ of the 
randomly occupied bonds and $\chi (x ,x^\prime)$ is an indicator function which is is unity 
if $x$ and $x^\prime$ are connected and zero otherwise. $C_R^{(n)}$ is related to
the $2n$th multifractal moment of the current distribution via Cohn's theorem\cite{cohn_50} 
(cf.\cite{park_harris_lubensky_87}),
\begin{eqnarray}
\label{finalCumulant}
C_R^{(n)}(x ,x^\prime) = v_n \left\langle \chi (x ,x^\prime) \sum_b 
\left( I_b / I \right)^{2n} \right\rangle_C / \left\langle \chi (x ,x^\prime) 
\right\rangle_C \ , 
\end{eqnarray}
where $v_n = \left\{ \left( \delta \rho_b \right)^n \right\}_f^{(c)}$ is the $n$th cumulant 
of the deviation $\delta \rho_b = \rho_b - \overline{\rho}$ of the bond resistance $\rho_b = 
\sigma_b^{-1}$ from its average $\overline{\rho}$.

Our aim is to determine $C_R^{(n)}$. Hence, the task is to solve the set of 
Kirchhoff's equations and to perform the averages over the diluted lattice configurations
and the noise. It can be achieved by employing the replica technique\cite{stephen_78}. 
PHL introduced $D \hspace{-1mm}\times \hspace{-1mm}E$-fold replicated voltages $V_x \to 
\stackrel{\mbox{{\tiny $\leftrightarrow$}}}{V}_x = \left( V_x^{(\alpha ,\beta)} 
\right)_{\alpha ,\beta = 1}^{D,E}$ and $\psi_{\stackrel{\mbox{{\tiny 
$\leftrightarrow$}}}{\lambda}}(x) = \exp \left( i \stackrel{\mbox{{\tiny 
$\leftrightarrow$}}}{\lambda} \cdot \stackrel{\mbox{{\tiny $\leftrightarrow$}}}{V}_x 
\right)$, 
where $\stackrel{\mbox{{\tiny $\leftrightarrow$}}}{\lambda} \cdot \stackrel{\mbox{{\tiny 
$\leftrightarrow$}}}{V}_x = \sum_{\alpha , \beta =1}^{D,E} \lambda^{(\alpha , \beta )} 
V_x^{(\alpha , \beta)}$ and $\stackrel{\mbox{{\tiny $\leftrightarrow$}}}{\lambda} \neq 
\stackrel{\mbox{{\tiny $\leftrightarrow$}}}{0}$. The corresponding correlation functions 
are defined as
\begin{eqnarray}
\label{noisyErzeugendeFunktion}
\lefteqn{ G \left( x, x^\prime ;\stackrel{\mbox{{\tiny $\leftrightarrow$}}}{\lambda} 
\right) =
\lim_{D \to 0} \Bigg\langle \Bigg\{ \frac{1}{\prod_{\beta =1}^E Z \left( \left\{ 
\sigma_b^{(\beta )} \right\} , C \right)^D } 
\int \prod_j \prod_{\alpha ,\beta=1}^{D,E} 
dV_j^{(\alpha , \beta )}
}
\nonumber \\
& & \times \exp \bigg[ -\frac{1}{2} P \left( \left\{  \stackrel{\mbox{{\tiny 
$\leftrightarrow$}}}{V} \right\} \right) + i \stackrel{\mbox{{\tiny 
$\leftrightarrow$}}}{\lambda} \cdot \left( \stackrel{\mbox{{\tiny 
$\leftrightarrow$}}}{V}_x  - \stackrel{\mbox{{\tiny $\leftrightarrow$}}}{V}_{x^\prime} 
\right) \bigg] \Bigg\}_f \Bigg\rangle_C \ ,
\end{eqnarray}
where $ P \left( \left\{  \stackrel{\mbox{{\tiny $\leftrightarrow$}}}{V} \right\} \right) = 
\sum_{\alpha ,\beta =1}^{D,E} \sum_{\langle i,j \rangle} 
\sigma_{i,j}^{(\beta )} \left( V_i^{(\alpha ,\beta)} - V_j^{(\alpha ,\beta )}\right)^2 $ is 
the power dissipated on the backbone and $Z$ is the usual normalization.

The integrations in Eq.~(\ref{noisyErzeugendeFunktion}) can be carried out by employing 
the saddle point method, which is exact in this case. The maximum of the integrand is 
determined by the solution of Kirchhoff's equations and thus 
\begin{eqnarray}
\label{noisyGenFkt}
G \left( x, x^\prime ;\stackrel{\mbox{{\tiny $\leftrightarrow$}}}{\lambda} \right) = 
\left\langle \prod_{\beta =1}^E \left\{ \exp \left[ - 
\frac{\vec{\lambda}^{(\beta )2}}{2} R^{(\beta )} \left( x,x^\prime 
\right) \right] \right\}_f \right\rangle_C \ .
\end{eqnarray}
On defining $K_l \left( \stackrel{\mbox{{\tiny $\leftrightarrow$}}}{\lambda} \right) = 
\sum_{\beta =1}^E \left[ \sum_{\alpha =1}^D \left( \lambda^{(\alpha ,\beta )} \right)^2 
\right]^l$ one obtains
\begin{eqnarray}
\label{cumulantGenFkt}
G \left( x, x^\prime ;\stackrel{\mbox{{\tiny $\leftrightarrow$}}}{\lambda} \right) = 
\left\langle  \exp \left[ \sum_{l=1}^\infty \frac{(-1/2)^l}{l!} K_l \left( 
\stackrel{\mbox{{\tiny $\leftrightarrow$}}}{\lambda} \right) \left\{ R 
\left( x,x^\prime \right)^l \right\}_f^{(c)} \right]  \right\rangle_C \ ,
\end{eqnarray}
i.e., $G$ represents a generating function for $C_R^{(n)}$.

To guarantee that the limit $\lim_{D \to 0}Z^{DE}$ is well defined one switches to voltage 
variables $\stackrel{\mbox{{\tiny $\leftrightarrow$}}}{\theta}$ and current variables 
$\stackrel{\mbox{{\tiny $\leftrightarrow$}}}{\lambda}$ taking discrete values on a $D 
\hspace{-1mm}\times \hspace{-1mm}E$-dimensional torus. Upon Fourier transformation in replica 
space one introduces the Potts spins\cite{Zia_Wallace_75} 
$\Phi_{\stackrel{\mbox{\tiny $\leftrightarrow$}}{\theta}} \left( x \right) = (2M)^{-DE} 
\sum_{\stackrel{\mbox{\tiny $\leftrightarrow$}}{\lambda} \neq 
\stackrel{\mbox{\tiny $\leftrightarrow$}}{0}} \exp \left( i \stackrel{\mbox{\tiny 
$\leftrightarrow$}}{\lambda} \cdot \stackrel{\mbox{\tiny $\leftrightarrow$}}{\theta} 
\right) \psi_{\stackrel{\mbox{\tiny $\leftrightarrow$}}{\lambda}} (x) = 
\delta_{\stackrel{\mbox{\tiny $\leftrightarrow$}}{\theta}, \stackrel{\mbox{\tiny 
$\leftrightarrow$}}{\theta}_{x}} - (2M)^{-DE} $ subject to the condition 
$\sum_{\stackrel{\mbox{\tiny $\leftrightarrow$}}{\theta }} 
\Phi_{\stackrel{\mbox{\tiny $\leftrightarrow$}}{\theta}} \left( x \right) = 0$. The effective 
Hamiltonian 
\begin{eqnarray}
H_{\mbox{\scriptsize{rep}}} =  - \ln \left\langle \left\{ \exp \left[ - \frac{1}{2} P \left( 
\left\{ \stackrel{\mbox{\tiny $\leftrightarrow$}}{\theta} \right\} \right) \right] \right\}_f 
\right\rangle_C
\end{eqnarray}
can be expanded in terms of the Potts spins as
\begin{eqnarray}
H_{\mbox{\scriptsize{rep}}} = - \sum_{\langle i ,j \rangle} \sum_{\stackrel{\mbox{\tiny 
$\leftrightarrow$}}{\theta},\stackrel{\mbox{\tiny $\leftrightarrow$}}{\theta}^\prime} K 
\left( \stackrel{\mbox{\tiny $\leftrightarrow$}}{\theta} - \stackrel{\mbox{\tiny 
$\leftrightarrow$}}{\theta}^\prime \right) \Phi_{\stackrel{\mbox{\tiny 
$\leftrightarrow$}}{\theta}} \left( i \right) \Phi_{\stackrel{\mbox{\tiny 
$\leftrightarrow$}}{\theta}^\prime} \left( j \right)
\end{eqnarray}
where
\begin{eqnarray}
K \left( \stackrel{\mbox{\tiny $\leftrightarrow$}}{\theta} \right) = \ln \left\{ 1 + 
\frac{p}{1-p} \exp \left[ \sum_{l=1}^\infty \frac{(-1/2)^l}{l!} \Delta^{(l)} K_l \left( 
\stackrel{\mbox{\tiny $\leftrightarrow$}}{\theta} \right) 
\right] \right\} \ .
\end{eqnarray} 
The Fourier transform $\widetilde{K} \big( \stackrel{\mbox{\tiny $\leftrightarrow$}}{\lambda} 
\big)$ of $K \big( \stackrel{\mbox{\tiny $\leftrightarrow$}}{\theta} \big)$
in turn has the expansion
\begin{eqnarray}
\label{noisyTaylorExp}
\widetilde{K} \left( \stackrel{\mbox{\tiny $\leftrightarrow$}}{\lambda} \right) = \tau + 
w \stackrel{\mbox{\tiny $\leftrightarrow$}}{\lambda}^2 + \sum_{l=2}^{\infty} v_l K_l 
\left( \stackrel{\mbox{\tiny $\leftrightarrow$}}{\lambda} \right) + \cdots \ ,
\end{eqnarray}
with $\tau$, $w \sim \overline{\sigma}^{-1}$, and $v_{l} \sim \Delta^{(l)} / 
\overline{\sigma}^{2l}$ being expansion coefficients. We display only the leading parts of 
the expansion. The terms proportional to $K_l 
\big( \stackrel{\mbox{\tiny $\leftrightarrow$}}{\lambda} \big)$ are retained because they are 
required in extracting the multifractal moments from $G \big( x, x^\prime 
;\stackrel{\mbox{{\tiny $\leftrightarrow$}}}{\lambda} \big)$.

We proceed with the usual coarse graining step and replace the Potts spins 
$\Phi_{\stackrel{\mbox{{\tiny $\leftrightarrow$}}}{\theta}} \left( x \right)$ by order 
parameter fields $\varphi \big( {\rm{\bf x}} ,\stackrel{\mbox{{\tiny 
$\leftrightarrow$}}}{\theta} \big)$ which inherit the constraint 
$\sum_{\stackrel{\mbox{{\tiny $\leftrightarrow$}}}{\theta}} \varphi \big( {\rm{\bf x}} 
,\stackrel{\mbox{{\tiny $\leftrightarrow$}}}{\theta} \big) = 0$. We model the 
corresponding field theoretic Hamiltonian $\mathcal H \mathnormal$ in the spirit 
of Landau as a mesoscopic free energy from local monomials of the order parameter field 
and its gradients in real and replica space. The gradient expansion is justified since 
the interaction is short ranged in both spaces for large $\overline{\sigma}$. Purely local 
terms in replica space have to respect the full $S_{\left( 2M \right)^{DE}}$ Potts symmetry. 
After these remarks we write down the Landau-Ginzburg-Wilson type Hamiltonian
\begin{eqnarray}
\label{noisyFinalHamil}
{\mathcal H \mathnormal} = \int d^dx {\textstyle \sum_{\stackrel{\mbox{{\tiny 
$\leftrightarrow$}}}{\theta}} }
\bigg\{  \frac{1}{2} \varphi \, K \left( \nabla ,
\nabla_{\stackrel{\mbox{{\tiny $\leftrightarrow$}}}{\theta}} \right) \varphi + 
\frac{g}{6}\varphi^3 \bigg\} \ ,
\end{eqnarray}
where
\begin{eqnarray}
K \left( \nabla ,\nabla_{\stackrel{\mbox{{\tiny $\leftrightarrow$}}}{\theta}} \right) = 
\tau + \nabla^2 + w \sum_{\alpha , \beta =1}^{D,E} \frac{- \partial^2}{\left( \partial 
\theta^{(\alpha ,\beta )} \right)^2 } 
+ \sum_{l=2}^\infty v_l \sum_{\beta=1}^E \left[ \sum_{\alpha =1}^D   
\frac{- \partial^2}{\left( \partial \theta^{(\alpha ,\beta )} \right)^2 } \right]^{l} \ .
\end{eqnarray}
In Eq.~(\ref{noisyFinalHamil}) we have neglected terms  which are irrelevant in the 
renormalization group sense. $\tau$, $w$ and $v_l$ are now coarse grained analogues of the 
original coefficients appearing in Eq.~(\ref{noisyTaylorExp}). Note that $\mathcal H 
\mathnormal$ reduces to the usual $(2M)^{DE}$ states Potts model Hamiltonian by setting $v_l 
=0$ and $w=0$ as one retrieves purely geometrical percolation in the limit of vanishing $v_l$ 
and $w$.

Now we analyze the relevance of the $v_l$. A straightforward scaling analysis reveals that
\begin{eqnarray}
\label{cumulantScaling}
C_R^{(l)} \left( \left( {\rm{\bf x}}, {\rm{\bf x}}^\prime \right) ; \tau , w, \left\{ 
v_l \right\} \right) = w^l f_l \left( \left( {\rm{\bf x}}, {\rm{\bf x}}^\prime \right) ; 
\tau , \left\{ \frac{v_k}{w^k} \right\} \right) \ ,
\end{eqnarray}
where $f_l$ is a scaling function. Note that the coupling constants $v_k$ appear only as $v_k 
/ w^k$. Dimensional analysis of the Hamiltonian shows that $w \stackrel{\mbox{{\tiny 
$\leftrightarrow$}}}{\lambda}^2 \sim \mu^2$ and  $v_k K_k \big( \stackrel{\mbox{{\tiny 
$\leftrightarrow$}}}{\lambda} \big) \sim \mu^2$, where $\mu$ is an inverse length scale. 
Thus, $v_k / w^k \sim \mu^{2-2k}$ and hence the $v_k / w^k$ have a negative naive dimension 
which decreases drastically with increasing $k$. This leads to the conclusion that the 
$v_{k}$ are highly irrelevant couplings. Though irrelevant, one must not set $v_{k}=0$ in 
calculating the noise exponents. We expand the scaling function $f_l$ yielding
\begin{eqnarray}
\label{expOfCumulantScaling1}
C_R^{(l)} \left( \left( {\rm{\bf x}}, {\rm{\bf x}}^\prime \right) ; \tau , w, \left\{ 
v_l 
\right\} \right) = v_l \left\{ C_l^{(l)} + C_{l+1}^{(l)} \frac{v_{l+1}}{w v_l} + \cdots 
\right\} \ ,
\end{eqnarray}
with $C_k^{(l)}$ being expansion coefficients depending on ${\rm{\bf x}}$, ${\rm{\bf 
x}}^\prime$, and $\tau$. It is important to recognize that $C_{k<l}^{(l)} = 0$ because 
the corresponding terms are not generated in the perturbation calculation. The first term on 
the right hand side of Eq.~(\ref{expOfCumulantScaling1}) gives the leading behavior. Thus, 
$C_R^{(l)}$ vanishes upon setting $v_l = 0$ and we cannot gain any further information about 
$C_R^{(l)}$. In other words, the $v_{l}$ are dangerously irrelevant in investigating the 
critical properties of the $C_R^{(l\geq 2)}$.

Our renormalization group improved perturbation calculation comprises the diagrams A to L given in Ref.\cite{janssen_stenull_99}. Since they are irrelevant, the couplings proportional to $v_l$ have to be treated by 
inserting\cite{amit_zinn-justin} 
\begin{eqnarray}
\label{opdev}
{\mathcal O \mathnormal}^{(l)} = - \frac{1}{2} v_l \int d^d p \, \, {\textstyle \sum_{\big\{ \stackrel{\mbox{{\tiny $\leftrightarrow$}}}{\lambda} 
\big\}} }K_l \left( \stackrel{\mbox{{\tiny 
$\leftrightarrow$}}}{\lambda} \right) \phi \left( {\rm{\bf p}} , 
\stackrel{\mbox{{\tiny 
$\leftrightarrow$}}}{\lambda} \right) \phi \left( -{\rm{\bf p}} , 
- \stackrel{\mbox{{\tiny 
$\leftrightarrow$}}}{\lambda} \right)
\end{eqnarray}
into these diagrams. Here $\phi \big( {\rm{\bf p}} , \stackrel{\mbox{{\tiny $\leftrightarrow$}}}{\lambda} 
\big)$ denotes the Fourier transform of $\varphi \big( {\rm{\bf x}} , 
\stackrel{\mbox{{\tiny $\leftrightarrow$}}}{\theta} \big)$. In Schwinger parametrization
the computation of the diagrams (including insertions) involves summations of the 
structure
\begin{eqnarray}
- s_i v_l   K_l \left( \stackrel{\mbox{{\tiny $\leftrightarrow$}}}{\lambda}_i \right) \exp 
\left[ -w  \sum_j s_j 
\stackrel{\mbox{{\tiny $\leftrightarrow$}}}{\lambda}_j^2 \right] \ .
\end{eqnarray}
We replace these summations by integrations which can be carried out by completing the squares in the exponential. In the limit $D \to 0$ we obtain
\begin{eqnarray}
\label{huhu}
- s_i c_i \left( \left\{ s \right\} \right)^{2l} v_l K_l \left( \stackrel{\mbox{{\tiny 
$\leftrightarrow$}}}{\lambda} \right) + \cdots \ , 
\end{eqnarray}
where the $c_i \left( \left\{ s \right\} \right)$ are homogeneous functions of the Schwinger paramters $s_j$ of degree zero. After all, the total contribution of a diagram can be written as
\begin{eqnarray}
\label{noisyExpansionOfDiagrams}
I \left( {\rm{\bf p}}^2 , \stackrel{\mbox{{\tiny $\leftrightarrow$}}}{\lambda} \right) 
= - v_l K_l \left( \stackrel{\mbox{{\tiny $\leftrightarrow$}}}{\lambda} \right) \int_0^\infty 
\prod_i ds_i \, \sum_j s_j \, c_j \left( \left\{ s \right\} \right)^{2l}
 D \left( {\rm{\bf p}}^2, \left\{ 
s \right\} \right) + \cdots \ .
\end{eqnarray}
Here $D \left( {\rm{\bf p}}^2, \left\{ s \right\} \right)$ stands for the integrand 
one obtains upon Schwinger parametrization of the corresponding diagram in the usual 
$\phi^3$ theory.

The ellipsis in Eq.~(\ref{noisyExpansionOfDiagrams}) stands for primitive divergences  
corresponding to all operators ${\mathcal O \mathnormal}^{(l)}_i$ of the generic form 
$\stackrel{\mbox{{\tiny $\leftrightarrow$}}}{\lambda}^{2a} \hspace{-2mm} {\rm{\bf p}}^{2b} 
\phi^n$ having the same or a lower naive dimension than ${\mathcal O \mathnormal}^{(l)}$ 
($a+b+n \leq l+2$). This myriad of newly generated operators is required as counterterms in 
the Hamiltonian. Inserting either of the ${\mathcal O \mathnormal}^{(l)}_i$, however, does 
not generate ${\mathcal O \mathnormal}^{(l)}$. Thus, the renormalization scheme is given by
\begin{eqnarray}
{\mathcal O \mathnormal}^{(l)}_{\mbox{\scriptsize bare \normalsize}} = Z^{(l)} \, {\mathcal O 
\mathnormal}^{(l)}_{\mbox{\scriptsize ren \normalsize}} + \sum_i Z^{(l)}_i \, {\mathcal O 
\mathnormal}^{(l)}_{i, \hspace{1mm}\mbox{\scriptsize ren \normalsize}}
\\
{\mathcal O \mathnormal}^{(l)}_{i, \hspace{1mm}\mbox{\scriptsize bare \normalsize}} = \sum_i 
Z^{(l)}_{i,j} \,  {\mathcal O \mathnormal}^{(l)}_{j, \hspace{1mm}\mbox{\scriptsize ren 
\normalsize}} \ ,
\end{eqnarray}
and one solely needs $Z^{(l)}$ in calculating the scaling index of ${\mathcal O 
\mathnormal}^{(l)}_{\mbox{\scriptsize ren \normalsize}}$. Therefore, we refer to ${\mathcal O 
\mathnormal}^{(l)}_{\mbox{\scriptsize ren \normalsize}}$ as master and to the ${\mathcal O 
\mathnormal}^{(l)}_{i, \hspace{1mm}\mbox{\scriptsize ren \normalsize}}$ as his servants.

\begin{figure}
\epsfxsize=13cm
\centerline{\epsffile{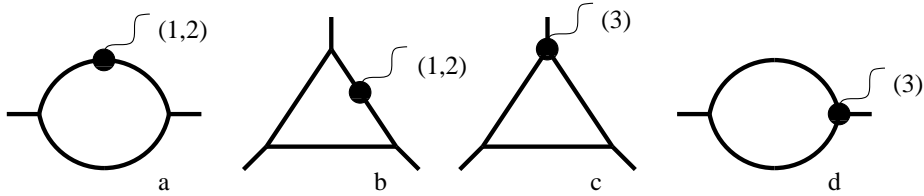}}
\caption[]{\label{fig1}One-loop diagrams obtained by inserting (1) ${\mathcal O \mathnormal}^{(2)}_1$, (2) ${\mathcal O \mathnormal}^{(2)}_2$ and (3) ${\mathcal O \mathnormal}^{(2)}_3$ respectively.}
\end{figure}

Now we briefly illustrate our perturbation calculation at one-loop level. Upon inserting ${\mathcal O \mathnormal}^{(2)}$ into the conducting propagators of diagrams A and B one finds in $\epsilon$-expansion
\begin{eqnarray}
\label{exA1}
\mbox{A}_{{\mathcal O \mathnormal}^{(2)}} - 2 \, \mbox{B}_{{\mathcal O \mathnormal}^{(2)}}&=&  g^2 \frac{G_\epsilon}{\epsilon} \tau^{-\epsilon /2} \Bigg\{ \frac{14}{15} v_2 
K_2 \left( \stackrel{\mbox{{\tiny $\leftrightarrow$}}}{\lambda} \right) + \frac{1}{15} v_2 \left( \stackrel{\mbox{{\tiny $\leftrightarrow$}}}{\lambda}^2 \right)^2 + \frac{1}{15} 
\frac{v_2}{w}  \stackrel{\mbox{{\tiny $\leftrightarrow$}}}{\lambda}^2 {\rm{\bf p}}^2  \Bigg\} \ ,
\end{eqnarray}
where $G_\epsilon = \left( 4 \pi \right)^{(-d/2)} \Gamma \left( 1 + \epsilon /2 \right)$. We learn, that not only primitive divergencies proportional to $K_2 \left( \stackrel{\mbox{{\tiny $\leftrightarrow$}}}{\lambda} \right)$, but also proportional to ${\rm{\bf p}}^2 \stackrel{\mbox{{\tiny $\leftrightarrow$}}}{\lambda}^2$ and $\left( \stackrel{\mbox{{\tiny $\leftrightarrow$}}}{\lambda}^2 \right)^2$ are generated. Thus, one has, at least in principle, to consider ${\mathcal O \mathnormal}^{(2)}$ in liaison with
\begin{eqnarray}
{\mathcal O \mathnormal}^{(2)}_1 &=& - \frac{1}{2} v_2 \int d^d p \, \, {\textstyle \sum_{\big\{ \stackrel{\mbox{{\tiny $\leftrightarrow$}}}{\kappa} 
\big\}} } \left( \stackrel{\mbox{{\tiny 
$\leftrightarrow$}}}{\lambda}^2 \right)^2 \phi \left( {\rm{\bf p}} , 
\stackrel{\mbox{{\tiny 
$\leftrightarrow$}}}{\lambda} \right) \phi \left( -{\rm{\bf p}} , 
- \stackrel{\mbox{{\tiny 
$\leftrightarrow$}}}{\lambda} \right) \ ,
\nonumber \\
{\mathcal O \mathnormal}^{(2)}_2 &=& - \frac{1}{2} \frac{v_2}{w} \int d^d p \, \, {\textstyle \sum_{\big\{ \stackrel{\mbox{{\tiny $\leftrightarrow$}}}{\kappa} 
\big\}} }  \stackrel{\mbox{{\tiny 
$\leftrightarrow$}}}{\lambda}^2 {\rm{\bf p}}^2 \phi \left( {\rm{\bf p}} , 
\stackrel{\mbox{{\tiny 
$\leftrightarrow$}}}{\lambda} \right) \phi \left( -{\rm{\bf p}} , 
- \stackrel{\mbox{{\tiny 
$\leftrightarrow$}}}{\lambda} \right) \ ,
\nonumber \\
{\mathcal O \mathnormal}^{(2)}_3 &=& - \frac{1}{18} \frac{v_2}{w} \int d^d p_1 d^d p_2 d^d p_3 \, \, {\textstyle \sum_{\big\{ \stackrel{\mbox{{\tiny $\leftrightarrow$}}}{\lambda}_1 , \stackrel{\mbox{{\tiny $\leftrightarrow$}}}{\lambda}_2 , \stackrel{\mbox{{\tiny $\leftrightarrow$}}}{\lambda}_3
\big\}} }  \sum_i \stackrel{\mbox{{\tiny 
$\leftrightarrow$}}}{\lambda}_i^2 \phi \left( {\rm{\bf p}}_1 , 
\stackrel{\mbox{{\tiny 
$\leftrightarrow$}}}{\lambda}_1 \right) \phi \left( {\rm{\bf p}}_2 , 
\stackrel{\mbox{{\tiny 
$\leftrightarrow$}}}{\lambda}_2 \right) \phi \left(  {\rm{\bf p}}_3 , 
\stackrel{\mbox{{\tiny 
$\leftrightarrow$}}}{\lambda}_3 \right) \ ,
\end{eqnarray}
where $\sum_i \stackrel{\mbox{{\tiny $\leftrightarrow$}}}{\lambda}_i = \stackrel{\mbox{{\tiny $\leftrightarrow$}}}{0}$ and $\sum_i {\rm{\bf p}}_i = {\rm{\bf 0}}$. The diagrams one obtains upon insertion of these operators are depicted in Fig.~\ref{fig1}. A straightforward calculation reveals that neither of the diagrams in Fig.~\ref{fig1} contains primitive divergencies proportional to $K_2 \left( \stackrel{\mbox{{\tiny $\leftrightarrow$}}}{\lambda} \right)$. Hence, these diagrams can be neglected in calculating the scaling index of ${\mathcal O \mathnormal}^{(2)}$.

The situation is different if one wishes to compute corrections to scaling associated with ${\mathcal O \mathnormal}^{(2)}_1$\cite{stenull_janssen_2000}. All the diagrams in Fig.~\ref{fig1} enter the correction exponent and thus one has to compute and diagonalize a $3\times 3$ renormalization matrix. This was overlooked by Harris and Lubensky\cite{Harris_Lubensky_87} who erroneously neglected diagrams b, c, d and diagram a with the ${\mathcal O \mathnormal}^{(2)}_2$ insertion  .

By employing dimensional regularization and minimal subtraction we proceed with standard 
techniques of renormalized field theory\cite{amit_zinn-justin}. We calculate $Z^{(l)}$ to 
two-loop order. From the renormalization group equation and dimensional analysis we deduce 
that the correlation function $G$ scales at criticality as
\begin{eqnarray}
G \left( {\bf x}, {\bf x}^\prime ;\stackrel{\mbox{{\tiny $\leftrightarrow$}}}{\lambda} 
\right) = |{\bf x}-{\bf x}^\prime 
|^{2-d-\eta} \Big\{ 1 + w \stackrel{\mbox{{\tiny $\leftrightarrow$}}}{\lambda}^2 |{\bf 
x}-{\bf x}^\prime |^{\phi/\nu} 
+ v_l K_l \left( \stackrel{\mbox{{\tiny $\leftrightarrow$}}}{\lambda} \right) |{\bf x}-{\bf 
x}^\prime |^{\psi_l/\nu} + \cdots \Big\} \ .
\end{eqnarray}
$\nu$ and $\eta$ are the well known critical exponents for percolation\cite{alcantara_80}. 
$\phi$ is the resistance exponent\cite{lubensky_wang_85,stenull_janssen_oerding_99}, 
$\phi=1+\epsilon /42 +4\epsilon^2 /3087 + {\em O} \left( \epsilon^3  \right)$. For the noise 
exponents $\psi_l$, $l\geq 2$, we obtain here
\begin{eqnarray}
\label{monsterExponent}
\psi_l &=& 1 + \frac{\epsilon}{7  \left( 1+l \right) \left( 1+2l \right)} + 
\frac{\epsilon^2}{12348
      \left( 1 + l \right)^3
      \left( 1 + 2l \right)^3}
\nonumber \\
&\times& \Big\{
 313 - 672\gamma + 
        l\Big\{ 3327 - 4032\gamma - 
           8l\Big\{ 4
               \left( -389 + 273\gamma \right)
\nonumber \\               
&+&                   
              l\left[ -2076 + 1008\gamma + 
               l  \left( -881 + 336\gamma
                     \right) \right]  \Big\}  \Big\}
\nonumber \\                      
            &-& 
        672\left( 1 + l \right)^2
         \left( 1 + 2l \right)^2
         \Psi(1 + 2l)  \Big\} + {\sl O} \left( \epsilon^3  \right) \ .
\end{eqnarray}
$\gamma =0.5772...$ denotes Euler's constant and $\Psi$ stands for the Digamma function. With 
Eq.~(\ref{cumulantGenFkt}) the desired scaling behavior of $C_R^{(n)}$ is now readily derived 
yielding 
\begin{eqnarray}
C_R^{(n)} \sim |{\bf x}-{\bf x}^\prime |^{\psi_n/\nu} \ .
\end{eqnarray}

Our result for the noise exponents is in agreement to first order in $\epsilon$ with the 
one-loop calculation by PHL. We point out that Eq.~(\ref{monsterExponent}) can be 
analytically continued to $l=1$ and is in conformity with the result for $\phi$ cited above. 
Analytic continuation of $\psi_l$ to $l=0$ and comparison with the available 
$\epsilon$-expansion results for 
$D_B$\cite{harris_lubensky_83,janssen_stenull_oerding_99,janssen_stenull_99} shows that 
$\psi_{0} = \nu D_B$\cite{arcangelis_etal_85} up to order ${\em O} \left( \epsilon^3 
\right)$. Blumenfeld {\em et al.}\cite{blumenfeld_etal_87} proved that $\psi_l$ is a convex 
monotonically decreasing 
function of $l$. Note that our result for $\psi_l$ captures this feature for reasonable 
values of $\epsilon$. It reduces to unity in the limit $l\to \infty$ as one expects from the 
relation of $\psi_\infty$ to the fractal dimension of the singly connected (red) 
bonds\cite{arcangelis_etal_85}, $\psi_\infty = d_{\mbox{{\scriptsize red}}} \nu$, and 
Coniglio's proof\cite{coniglio_81_82} of $d_{\mbox{{\scriptsize red}}} = 1/\nu$. 

In conclusion, we introduced the concept of master and servant operators and showed that 
it works consistently as a tool to describe the multifractal properties of RRN by renormalized field theory. We presented the premier two-loop calculation of a family of multifractal exponents. Our result (\ref{monsterExponent}) is for dimensions near the upper critical dimension 6 the most accurate analytic estimate of $\left\{ \psi_l \right\}$ that we know of. It fulfills several consistency checks.

We showed a one-to-one correspondence of the multifractal moments and the master 
operators. Though a myriad of servant operators is involved in the renormalization of 
the masters ${\mathcal O \mathnormal}^{(l)}$ the scaling behavior of the $l$th 
multifractal moment is governed by ${\mathcal O \mathnormal}^{(l)}$ only. The situation 
is different for operators which are irrelevant without being masters. The scaling behavior of the related quantities is influenced by the whole bunch of operators generated in the perturbation calculation. 

The concept of master and servant operators should have many more applications. Indeed, 
its applicability might be a general feature of multifractal systems. Hence, our concept 
could prove to be a key in understanding multifractality, at least from the standpoint of 
renormalized field theory.

We acknowledge support by the Sonderforschungsbereich 237 ``Unordnung und 
gro{\ss}e Fluktuationen'' of the Deutsche Forschungsgemeinschaft. 


\end{document}